\documentclass[nofootinbib]{revtex4-1}
\usepackage{silence}
\usepackage{bm}
\usepackage{hyperref}
\usepackage{amsmath,amssymb,amsfonts,graphicx,slashed,bbm}
\usepackage[paper=letterpaper,margin=1in]{geometry}
\usepackage{braket, color}
\usepackage{upgreek}

\newcommand{\beq}{\begin{eqnarray}}
\newcommand{\eeq}{\end{eqnarray}}
\usepackage{amsmath}
\usepackage{amssymb}
\usepackage[paper=letterpaper,margin=1in]{geometry}
\usepackage{braket}
\usepackage{upgreek}
\usepackage{tikz}

\begin{document}

\title{ Not Just a Phase:  Bose Metal Unveiled}
\author{Philip W. Phillips}

\affiliation{Department of Physics and Institute for Condensed Matter Theory,
University of Illinois
1110 W. Green Street, Urbana, IL 61801, U.S.A.}

\date{\today}

\begin{abstract}
I analyze the recent observation of a Bose metal in NbSe$_2$ and place these results in the wider context of the first sighting of such a phase in amorphous Bi and MoGe thin films.

\end{abstract}
\maketitle

Whether they as elemental as photons or as complex as helium atoms, bosons exhibit the rather democratic property of being able to occupy the same quantum state irrespective of filling.  Consequently, the fermionic concept of partial occupancy of a quantum state is lacking for bosons and as a result, so is the associated rule that partially filled bands produce metallic transport.  However, experiments by Adam Tsen and colleagues\cite{Tsen2015}, reported in {\it Nature Physics} suggest that
a Bose metallic state can form once a two-dimensional superconductor is destroyed.  

\begin{figure}[t]
\includegraphics[width=5cm]{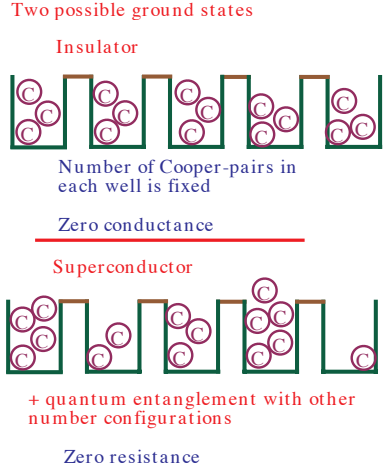}
\caption{Insulating and superconducting ground states of bosons illustrated here for Cooper 
(C) pairs.  At play here is the conjugacy between phase and number 
fluctuations of the Cooper pairs. In the insulator, Cooper pair 
number fluctuations cease leading to infinite uncertainty in the 
phase. Contrastly, in a superconductor, phase coherence obtains, 
leading thereby to infinite uncertainty in the Cooper pair 
particle number.} 
\label{fig1}
\end{figure}

One reason why a metallic state for bosons is {\it a priori} problematic  is that bosons are typically known\cite{fisher} to exist in one of two ground states that arise from the conjugacy between particle number and phase as illustrated in Fig. (\ref{fig1}).  The eigenstate of phase is a superfluid (superconductor if the bosons are charged) in which the particle number is completely indeterminate.  By contrast, the state in which the particle number is definite (indeterminate phase) does not permit the transport of bosons and hence is an insulator.    The Bose metal is then a state of matter in which large phase fluctuations do not necessarily kill the particle number fluctuations that are necessary to produce a conducting state intermediate between a superconductor and an insulator.

 Tsen, et al.\cite{Tsen2015} have observed this phenomenon in none other than NbSe$_2$, a material in which superconductivity in an atomically thin layer was first observed in 2009\cite{liu2009} albeit with a reduced transition temperature of $T_c\approx 2.5K$ compared with the bulk value of $T_c=7.2K$.  The reduction in $T_c$ is due to the enhanced role phase fluctuations play in 2D where superconductivity is destroyed by vortex-antivortex unbinding\cite{halperin}.  From the resistivity, Tsen, et al.\cite{Tsen2015} extract an onset temperature of $T_c\approx 5.2K$ which is slightly higher than the $T_c\approx 5.01K$ extracted from the temperature at which the voltage-current characteristics obey the non-linear relationship $V\approx I^3$ as is expected from the  vortex-antivortex unbinding model\cite{halperin}.  This agreement corroborates that the observed superconductivity is really in the 2D limit in which the thickness is less than the Cooper pair coherence length.  
 
 The key surprise here is that in addition to exhibiting a true zero resistance state for magnetic fields less than $0.175T$, Tsen, et al.\cite{Tsen2015} find that the resistivity levels at low temperatures for a wide parameter range,  $0.175T<H<3T$. The upper field here corresponds to the onset of the normal state in which Cooper pairs are no longer the charge carriers.  Hence, the claim is that for magnetic fields within $0.175T<H<3T$, charge $2e$ Cooper pairs are the charge carriers and they neither condense nor insulate but are characterized by a finite resistivity at $T=0$.  That is, they constitute a true Bose metal phase!   

The debate over the existence of a Bose metal is long-standing\cite{G1989,ephron,yazdani,mason,G2005,yoon}, but these data add three key facts to the discussion.
First, the Bose metal is observed in clean crystalline samples, thereby ruling out crystalline disorder as a root cause of the effect.  Second, the resistivity curves in NbSe$_2$ below 3T all display a thermally activated temperature dependence indicative of activated\cite{larkin} vortex-antivortex motion.  As in the thin film work previously\cite{yazdani,ephron,yoon}, the temperature at which the activated behavior ceases and the leveling of the resistivity obtains decreases as the field increases .  If the Bose metal were due to a heating effect, then the failure to cool the sample would occur at the same temperature, not one determined by the magnetic field.  Consequently, the observation here, as well as in previous samples,\cite{ephron,yazdani,yoon,G1989} seems profoundly robust.  Third, once the thermally activated flux-flow regime terminates, the state that obtains has a resistivity that increases as a power of the distance from the zero-resistance state. 

What does the Bose metal state look like?  Historically, the standard approach\cite{spivakkiv,kivchak,otterlo}was to include some extra degree of freedom that introduces dissipation to possibly generate a phase intermediate between those shown in Fig. (\ref{fig1}).  Regardless of the model chosen, the crucial metric for transport is the conductivity.  However, even for the simplest model describing the phase transition between the two ground states in Fig. (\ref{fig1}), there was no consensus\cite{girvin,otterlo,girvin2} as to how to compute the conductivity until Damle and Sachdev\cite{DS1997} showed that close to the transition region, the conductivity is a universal function of the form $\sigma_Q f(\omega/T)$ where $f$ is a monotonically decreasing function of the frequency, $\omega$, and the temperature, $T$.  The experiments correspond to the limiting procedure $\lim_{T\rightarrow 0}\lim_{\omega\rightarrow 0}$, that is to $f(0)$ not the inverse limit where $f(\infty)$ enters. The physics of $f(0)$ is pure hydrodynamics in which it is collisions of the quasiparticle excitations of the order parameter that regularize the conductivity.  The first implementation of the Damle/Sachdev\cite{DS1997} hydrodynamic approach to the conductivity\cite{dalidovich} in the standard dissipative models adopted for the insulator-superconductor transition showed that although dissipation can drive an intermediate region in temperature where the resistivity levels, ultimately at $T=0$ a superconductor obtains.  Hence, dissipation alone cannot drive the Bose metal.   Of course, appealing to single electrons to provide a $T=0$ metallic state in 2D is a non-starter because in this limit, disorder precludes\cite{go4} a metallic state. 

In light of these difficulties, it has been suggested\cite{CN} that weak disorder drives a new diffusive critical point in which the interactions enhance the conductivity, giving rise to a perfect conductor once superconductivity is destroyed.  This result highly constrains the correct theoretical description for the destruction of superconductivity in disordered interacting systems.  Namely, the phase that obtains once superconductivity is destroyed must have a resistivity that increases as a power law from the zero-resistance state.   

Of the theoretical models available, nicely reviewed by Tsen, et al.\cite{Tsen2015}, only two proposals predict a power-law growth\cite{DD,dp1,dp2,WP}  of the resistivity of the form $\rho(T=0)\propto \rho_c(g-g_c)^\alpha$.   The first\cite{DD} is a phenomenological one in which the finite resistivity  is due either to 1) unbinding  of dislocation-antidislocation pairs or 2) to free dislocations in a dirty system.  Since no explicit calculation is given\cite{DD}, it is unclear why the traditional vortex-antivortex activated behavior does not obtain considering that the temperature is much less than the binding energy.  In the second\cite{dp1,dp2,WP}, the conductivity is calculated for bosons moving in a glassy landscape arising from the randomness in the Josephson tunneling amplitudes.  Not only does the phase stiffness associated with superconductivity vanish in this model\cite{WP} but also the conductivity remains finite because bosonic excitations moving in an environment with many false minima, glassy dynamics, take an exponentially long time to find the ground state.  What distinguishes this model from the dissipative models used earlier to destroy phase coherence in 2D superconductors is the glassy environment.   

This work was extended\cite{WP} to the magnetic field-tuned transition and the resistivity in the Bose metal regime is directly proportional to the glassy order parameter and turns on as a power law, with a product of the exponents related to the temporal (dynamical) and spatial correlations.

Although a measurement of the turn-on exponent by Tsen et al.\cite{Tsen2015} is significant, as it is in agreement with theoretical predictions\cite{dp1,dp2,WP}, a true unveiling of the Bose metal would require a direct measurement of the dynamical and correlation exponents, $z$ and $\nu$ directly.  The  challenge is set.

\textbf{ Acknowledgements} I thank A. Yazdani, A. Pasupathy and S. Kivelson for helpful comments on earlier drafts.

\end{document}